\documentclass[twocolumn,iop,revtex4,usenatbib,twocolappendix]{openjournal}
\usepackage[T1]{fontenc}

\usepackage[english, english]{babel}

\DeclareRobustCommand{\VAN}[3]{#2}
\let\VANthebibliography\thebibliography
\def\thebibliography{\DeclareRobustCommand{\VAN}[3]{##3}\VANthebibliography}

\usepackage{algorithm} 
\usepackage[noend]{algpseudocode} 

\usepackage[bitstream-charter]{mathdesign}
\usepackage[usenames]{color}
\usepackage{xcolor}
\usepackage{soul}
\usepackage{multirow}
\usepackage{graphicx}
\usepackage{amsmath}
\usepackage{nicefrac}
\usepackage{microtype}
\usepackage{amssymb}
\usepackage{bbding}
\usepackage{csquotes} 
\usepackage{float}
\usepackage{xurl}
\usepackage{adjustbox}
\usepackage{comment}
\usepackage[hyperfootnotes=false]{hyperref}
\usepackage{tensor} 
\usepackage{enumerate}
\usepackage{orcidlink} 
\usepackage{booktabs} 

\hypersetup{colorlinks=true, citecolor=blue, linkcolor=blue, urlcolor=blue}
\definecolor{hotpink}{RGB}{255, 105, 180}

\definecolor{orcidlogocol}{HTML}{A6CE39}

\DeclareSymbolFont{usualmathcal}{OMS}{cmsy}{m}{n}
\DeclareSymbolFontAlphabet{\mathcal}{usualmathcal}

\definecolor{teal}{RGB}{0,128,128}

\newcommand{\ltsima}{$\; \buildrel < \over \sim \;$}
\newcommand{\lsim}{\lower.5ex\hbox{\ltsima}}
\newcommand{\gtsima}{$\; \buildrel > \over \sim \;$}
\newcommand{\gsim}{\lower.5ex\hbox{\gtsima}}
\newcommand{\bra}{\langle}
\newcommand{\ket}{\rangle}

\newcommand{\dd}{\mathrm{d}}

\newcommand{\likeli}{\mathcal{L}}

\newcommand{\pp}{\partial}

\algdef{SE}[REPEAT]{Repeat}{EndRepeat}[1]{\textbf{repeat} #1 \textbf{times}}{}

\begin{document}
\title{Approximating non-Gaussian Bayesian partitions with normalising flows: \\ statistics, inference and application to cosmology\vspace{-30pt}}

\author{Tobias R{\"o}spel\,\orcidlink{0009-0003-8645-4643}$^{1, \star}$}
\author{Adrian Schlosser\,\orcidlink{0009-0009-7118-9705}$^{1, \star}$}
\author{Bj{\"o}rn Malte Sch{\"a}fer\,\orcidlink{0000-0002-9453-5772}$^{1,2 \sharp}$}
\thanks{$^\star$ Both authors contributed equally to this work.}
\thanks{$^\sharp$ \href{mailto:bjoern.malte.schaefer@uni-heidelberg.de}{bjoern.malte.schaefer@uni-heidelberg.de}}

\affiliation{$^{1}$ Zentrum f{\"u}r Astronomie der Universit{\"a}t Heidelberg, Astronomisches Rechen-Institut, Philosophenweg 12, 69120 Heidelberg, Germany}
\affiliation{$^{2}$ Interdisziplinäres Zentrum f{\"u}r wissenschaftliches Rechnen der Universit{\"a}t Heidelberg, Im Neuenheimer Feld 205, 69120 Heidelberg, Germany}

\begin{abstract}
Subject of this paper is the simplification of Markov chain Monte Carlo sampling as used in Bayesian statistical inference by means of normalising flows, a machine learning method which is able to construct an invertible and differentiable transformation between Gaussian and non-Gaussian random distributions. We use normalising flows to compute Bayesian partition functions for non-Gaussian distributions and show how normalising flows can be employed in finding analytical expressions for posterior distributions beyond the Gaussian limit. Flows offer advantages for the numerical evaluation of the partition function itself, as well as for cumulants and for the information entropy. We demonstrate how normalising flows in conjunction with Bayes partitions can be used in inference problems in cosmology and apply them to the posterior distribution for the matter density $\Omega_m$ and a dark energy equation of state parameter $w_0$ on the basis of supernova data.
\end{abstract}


\maketitle

\section{Introduction}
Bayes'\@ theorem \citep[for reviews in its application to cosmology, see][]{trotta_bayes_2008, trotta_bayesian_2017} assembles the posterior distribution $p(\theta|y)$ of model parameters $\theta$ given an observation $y$ from the prior information $\pi(\theta)$ with the likelihood $\likeli(y|\theta)$ as the distribution of the data points $y$ for a given parameter choice $\theta$:  
\begin{equation}
p(\theta|y) = \frac{\likeli(y|\theta)\pi(\theta)}{p(y)},
\end{equation}
where the posterior distribution is normalised by the Bayesian evidence,
\begin{equation}
p(y) = \int\dd^n\theta\:\likeli(y|\theta)\pi(\theta),
\end{equation}
which plays as well an important role in Bayesian model selection \citep{jenkins_power_2011, handley_quantifying_2019, trotta_forecasting_2007, liddle_present_2006, kerscher_model_2019, knuth_bayesian_2015, modelselection}. A generalisation of Bayesian evidence is given by the canonical partition function $Z[T, J]$,
\begin{align}
    \label{eq:canZ}
    Z[T, &J] =
    \int\dd^n\theta\:\left[\likeli(y|\theta)\pi(\theta)\:\exp(J_\gamma\theta^\gamma)\right]^{1/T} \\
    &=
    \frac{1}{\widetilde{\mathcal{N}}(T)} \int\dd^n\theta\:\exp\left(-\frac{1}{T}\left[\frac{\chi^2(y|\theta)}{2}+\phi(\theta) - J_\gamma \theta^\gamma \right] \right) \nonumber
\end{align}
which falls back on the Bayesian evidence $p(y)$ for $T= 1$ and $J = 0$. $\widetilde{\mathcal{N}}(T) = \left(\mathcal{N}_{\likeli} \mathcal{N}_\pi \right)^{1/T}$ denotes the normalisation of the likelihood and prior respectively. By differentiation of the logarithmic partition function $-T \ln Z $, or equivalently the Helmholtz-free energy $F(T,J_\gamma)$ with respect to $J_\gamma$, cumulants of the posterior distribution $p(\theta|y)$ are generated \citep[for applications to cosmology, see][]{10.1093/mnras/Rover, kuntz2023partition, herzog2023partition, partition_info, 2023MNRAS.523.2027R}. The derivative of $\ln Z$ generates automatically correctly normalised expectation values and mirrors the fundamental structure of Bayes'\@ theorem with the numerator being the derivative of the denominator. At the same time, the partition function suggests an analogy to statistical physics, which explains the generation of samples by a Markov chain Monte Carlo algorithm in terms of the thermal motion of a particle inside a potential determined by the logarithmic likelihood $\chi^2/2$. $T$ and $J_\gamma$ are parameters which allow control over the sampling process, which itself constitutes in the language of statistical physics a canonical ensemble.

There are many methods for computing Bayesian evidences, which is in general a numerically challenging task. Nested sampling \citep{Skilling_2006, Ashton_2022, feroz2009multinest, dynesty_2020} has found a widespread application in cosmology and is considered the numerical standard, but competing algorithms exist, for instance population Monte Carlo \citep{kilbinger_bayesian_2010}, normalising flow based methods \citep{2024arXiv240505969P}, or macrocanonical sampling \citep{herzog2023partition}.

\begin{figure*}[t]
    \centering
    \includegraphics[width=.80\textwidth]{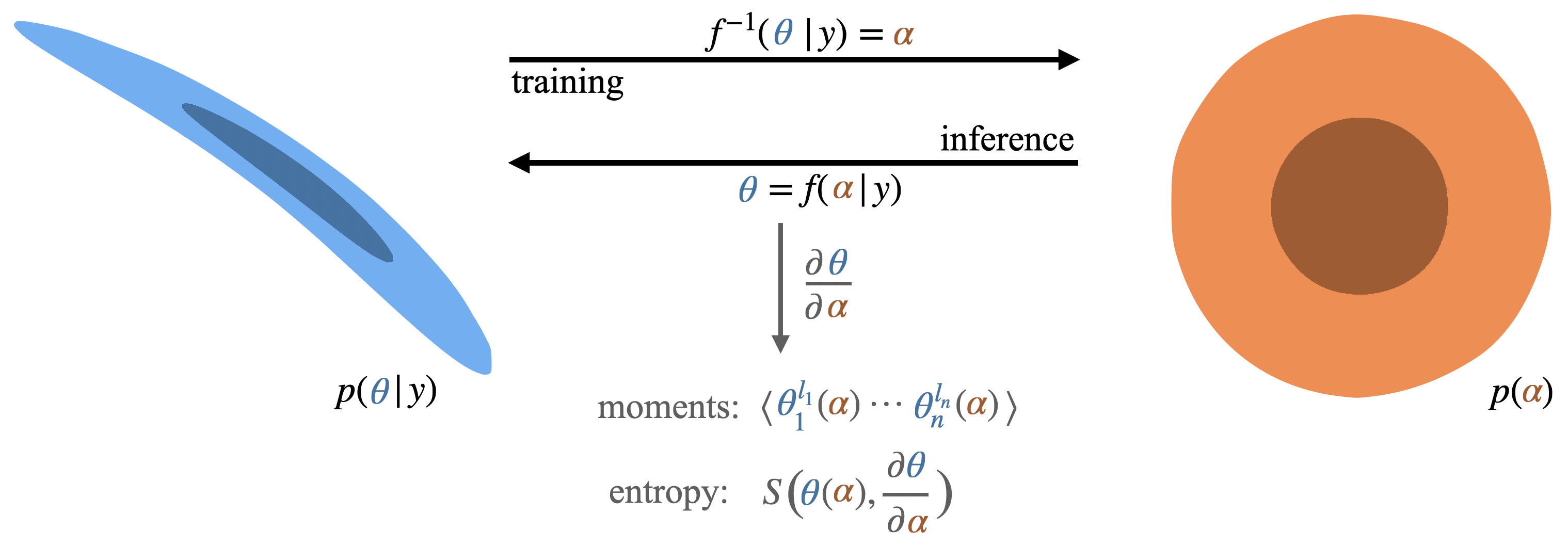}
    \caption{Schematic overview of the normalising flow learning the transformation $f^{-1}(\theta)$ from the posterior distribution $p(\theta|y)$ to a standard normal distribution $p(\alpha)$ for given data $y$ during training. Sampling from a standard normal distribution in $\alpha$ and using the inverted normalising flow $f(\alpha)$ allows to reconstruct the original posterior distribution $p(\theta|y)$. We use derivatives of also higher orders of the normalising flow to calculate the information entropy and moments of the posterior distribution.}
    \label{fig:overview}
\end{figure*}

Normalising flows \citep{2019arXiv191202762P, 2024arXiv240514392C, 2024arXiv240412294S, 2021arXiv210708001G} approach the issue of sampling from a non-Gaussian distribution: They construct a nonlinear, invertible mapping between a non-Gaussian and a Gaussian distribution, by minimisation of the Kullback-Leibler divergence. With this mapping, it is straightforward to generate samples from the non-Gaussian distribution and to estimate its properties such as moments or information entropies, or the Bayesian evidence itself, as demonstrated by \citet{2024arXiv240412294S} or with a slightly different focus by \cite{raveri2024understanding}. Normalising flows are recently becoming an important tool in cosmology and are used by \citet{mootoovaloo2024mathtt, 2024arXiv240505969P,2024arXiv240412294S} for example, in particular for the estimation of Bayesian evidences. \cite{harry3_prathaban2024accelerated} computes this quantity also within the context of temperature-dependent partition functions. Normalising flows are used for marginal Bayesian statistics in \cite{harry1_bevins2022marginal, harry2_bevins2023marginal}.

For our paper, we pursue the question of whether the differentiability of the mapping constructed by the normalising flow can be exploited, as many implementations enable auto-differentiability as a numerical method. This would be an alternative pathway to cumulants, entropies or the surprise statistic. After all, Bayes' partitions are generalisations of Bayes' evidence itself, so it would be sensible to expect that methods similar to those used for evidence computations should be applicable. In addition, we would like to find out whether the mapping constructed by the normalising flow can be integrated into analytical calculations in an advantageous way.

This paper is structured as follows: We discuss inference with non-linear models leading to non-Gaussian posterior distributions in Sect.~\ref{sect_flows} and demonstrate how Gaussianisations derived by normalising flows can be integrated into analytical calculations. We apply our methodology to the well-known non-Gaussian parameter space spanned by the matter density $\Omega_m$ and the dark energy equation of state parameter $w_0$ constrained by supernovae in Sect.~\ref{sect_cosmo}. Then, we demonstrate in Sects.~\ref{sect_geo} and~\ref{sect_temp_J} how the normalising flow modulates probability densities by introducing a nonlinear mapping and how it can be used to evaluate the partition functions in their dependence on temperature and control variables. Finally, we summarise our main results in Sect.~\ref{sect_summary} and defer technical details of the implementation to Appendix~\ref{sect_toy_model}. 

Throughout the paper, we adopt the summation convention and denote parameter tuples $\theta^\gamma$ and data tuples $y^i$ as vectors with contravariant indices; Greek indices are reserved for quantities in parameter space and Latin indices for data. For the cosmological application, we assume a flat, dark energy-dominated Friedmann-Lema{\^i}tre-Robertson-Walker cosmology with matter density $\Omega_m$ and a constant dark energy equation of state parameter $w_0$.
\vspace{0.6cm}

\section{Linking normalising flows to partition functions}\label{sect_flows}

\subsection{Normalising flows}
A normalising flow, first introduced in \citet{rezende2015variational}, is a neural network architecture that learns a map $f(\alpha) = \theta(\alpha)$ transforming standard normal distributed variables $\alpha$ to the parameters $\theta$ of an arbitrary distribution. Most importantly, this transformation $\alpha(\theta)\rightleftharpoons\theta(\alpha)$ is differentiable and invertible, i.e.\ a diffeomorphism. A common choice for the loss function is to minimise the Kullback-Leibler divergence \citep[for a Bayesian perspective]{baez_bayesian_2014} between a standard normal distribution $p(\alpha)$ and the distribution obtained by applying the transformation $\tilde{p} (\alpha) = | \det \mathrm{D} f(\alpha) |  p\left(\theta(\alpha)\right)$:
\begin{equation}
    \label{equ:dkl}
    D_{KL}(p(\alpha)|\tilde{p} (\alpha)) = \int\dd^n\alpha\: p(\alpha) \ln\left(\frac{p(\alpha)}{\tilde{p} (\alpha)}\right) \, .
\end{equation}
The loss function can be compactly written as
\begin{equation}
    \mathrm{Loss}(f) = \sum_{\text{samples} \,\, \alpha} \left( \frac{1}{2} \delta_{\rho \sigma} \alpha^{\rho} \alpha^{\sigma} - \ln | \det \mathrm{D} f(\alpha) | \right)    \, . 
\end{equation}
This is for example the standard suggestion in the \texttt{FrEIA} package \citep{freia} which is used for numerics in this paper. In summary, normalising flows allow generating samples of an arbitrary distribution $p(\theta)$ from samples of a standard normal distribution $p(\alpha)$ by learning the mapping $\theta = f(\alpha)$. The basic concept as well as how we perform calculations with the flow are illustrated by \autoref{fig:overview}.

\subsection{Normalising flows and partition functions}
Applying this to our partition function in \autoref{eq:canZ} leads to a Gaussianised partition function through change of variables, effectively through an integration by parts,
\begin{equation}
    \label{equ:Z_flow}
    Z[T,J] = 
    \frac{1}{\mathcal{N}(T)} \int\dd^n\alpha\: \exp\left(-\frac{1}{2 T}F_{\rho\sigma}\alpha^\rho\alpha^\sigma\right) g(\alpha)
\end{equation}
with
\begin{equation}
g(\alpha) = 
\left |\det\mathrm{D}f(\alpha)\right|^{1-\frac{1}{T}} \exp\left(\frac{J_\gamma \theta^\gamma(\alpha)}{T} \right) \,.
    \label{equ:g_alpha}
\end{equation}
In this case, the Fisher information matrix $F_{\rho\sigma} = \delta_{\rho\sigma}$ is the identity matrix for a standard uncorrelated normal distribution, and the normalisation is given by $\mathcal{N} (T) = ((2 \pi)^{n/2}/p(y))^{1/T}$.

This mapping replaces sampling from any physical, possibly non-Gaussian distribution with random variables $\theta$ by sampling from a standard normal distribution in $\alpha$. Important to note is the applicability of the change of variables, which allows to express the original distribution $p(\theta|y)$ in terms of the standard normal distribution $p(\alpha)$ as
\begin{align}
    \label{equ:change_of_variables}
	p(\alpha) = | \det \mathrm{D} f(\alpha|y) |  p(\theta(\alpha|y)),
\end{align}
where $\mathrm{D}f(\alpha|y) = \frac{\partial f}{\partial \alpha}$ is the Jacobian of $f (\alpha|y) = \theta(\alpha|y) $. Choosing a non-unit covariance is possible. A naive choice would be the actual Fisher-matrix, but as it originates from the unit matrix by a mere linear transform (given by the Cholesky decomposition), it is automatically taken care of by the normalising flow. In the following, we will omit for simplicity the dependence of the trained map $f$ on the data $y$ and thus just write $f(\alpha)$ instead of $f (\alpha|y)$.

\subsection{Moment and entropy calculations with a flow}\label{sect_momt_ent_calc}
Expectation values of an arbitrary function $A (\theta)$ play an important role in science. For example, one could be interested in $ A (\theta) = - \ln (p (\theta | y))$, which is the information entropy of the posterior $p (\theta | y)$. The expectation value of $A (\theta)$ is given by
\begin{equation}
    \langle A \rangle = \int \dd^n\theta\: p (\theta | y) \: A (\theta) \,.
\end{equation}
Using \autoref{equ:change_of_variables}, we can calculate the expectation value with respect to the standard normal distribution in $\alpha$ as
\begin{equation}
    \langle A \rangle = \int \dd^n \alpha\: p (\alpha) A (\theta (\alpha)) \,.
\end{equation}
Here, no determinant shows up as the integral and the probability density transform inversely to each other. If $A (\theta (\alpha))$ includes the probability density, special care is needed. It thus requires the change of variables formula and is not just plugging in $\theta = f (\alpha)$. Having samples $\{\alpha^i\}_N$ from a standard normal distribution, it is well known that the expectation value can be approximated by
\begin{equation}
    \label{equ:exp_val_approx}
    \langle A \rangle \approx \frac{1}{N} \sum_{i=1}^N \: A (\theta(\alpha^i)) \,.
\end{equation}
Furthermore, one can compute the moments of order $m$ by taking derivatives of the partition function with respect to the source terms $J$
\begin{align}\label{Moments}
	\langle \theta^{\gamma_1} \ldots \theta^{\gamma_n} \rangle
	= \left . \frac{1}{Z} \frac{\pp^m}{\pp{J_{\gamma_1}}\ldots\pp{J_{\gamma_m}}} Z \right |_{\substack{J=0\\ T = 1}}.
\end{align}
In the following chapters the skewness parameter $s^\gamma$ and kurtosis parameter $\kappa^\gamma$ of the posterior distribution $p(\theta^\gamma|y)$ will be of interest. They are simply defined as the third and fourth standardised moments, where the standardised moments are given by
\begin{align}\label{Standardized moments}
	\widetilde{\mu}_m = \left\langle \left (\frac{\theta^{\gamma}-\mu}{\sigma}\right )^m \right\rangle \,.
\end{align}
Here, $\mu$ and $\sigma$ correspond to the mean and variance of the corresponding marginal distribution $\Theta^\gamma$. The Helmholtz free energy can also be determined by the partition function and is given by
\begin{align}
    \label{equ:free_energy}
	F(T,J) = - T \ln Z[T,J]
\end{align}
Since $ \left . -\frac{\pp}{\pp T} F(T,J) \right |_{T=1, J = 0} = S = - \langle \ln p(\theta|y) \rangle$ one can also implicitly compute the entropy of the posterior distribution from the partition function $Z$. It is also possible to obtain the cumulants directly by differentiating $\ln Z$ instead of $Z$.

\subsection{Flow expansion through a differentiation operator}
Going further, one can use a well known formula from quantum field theory and apply it to the partition function in \autoref{equ:Z_flow}, which solves the integral by formulating it in terms of derivatives with respect to $\alpha$. Thus, \autoref{equ:Z_flow} becomes
\begin{align}
    \label{flow_expansion}
    Z[T, J] 
    &= \frac{1}{\mathcal{N}(T)} \int \mathrm{d}^n\alpha \: \exp \left(-\frac{1}{2 T} \delta_{\rho \sigma} \alpha^{\rho} \alpha^{\sigma} \right) g(\alpha) \nonumber \\
    &= \frac{ (2 \pi T)^{\frac{n}{2}}}{\mathcal{N}(T)} \exp \left( \frac{T}{2} \delta^{\rho \sigma} \frac{\partial}{\partial\alpha^\rho}\frac{\partial}{\partial\alpha^\sigma} \right) g(\alpha) \big|_{\alpha = 0} \,.
\end{align}
Here, $g(\alpha)$ is defined as in \autoref{equ:g_alpha}, and we will refer to this result as the flow expansion. At this point, we want to emphasise that solving the integral facilitates the analytical form and thus makes further statistical and respectively thermodynamic inspired calculations easier. In Sect. \ref{sect_cosmo}, we will comment on its numerical performance.

Again, interesting quantities are the moments of the posterior distribution $p(\theta | y)$. They can be determined by swapping the derivatives of $J$ with those taken by $\alpha$. For simplicity shown in one dimension, this reads
\begin{align}
	\langle \theta^m  \rangle
	&=  \left . \frac{1}{Z} \frac{\pp^m}{\pp{J^m}} Z[T,J] \right  |_{\substack{T=1 \\ J=0}}   \nonumber \\
	&= \exp \left( \frac{1}{2} \delta^{\rho \sigma} \frac{\partial}{\partial\alpha^\rho}\frac{\partial}{\partial\alpha^\sigma} \right) \theta(\alpha)^m g(\alpha) \big|_{\substack{\alpha = 0 \\ J=0}}    \nonumber \\
	\label{equ:laplace_1d}
	&= \left . \sum_{k=0}^{\infty} \frac{1}{2^k k!} \left(\frac{\partial}{\partial \alpha}\right)^{2k}  \theta(\alpha)^m \right |_{\alpha = 0}.
\end{align}
In the last step, the exponential function is replaced by its sum expression and $g(\alpha=0) = 1$ is inserted.\footnote{As a remark, we would like to note, that the operator-relation \citep{hermite_magic} $\text{He}_n(\alpha) = \exp\left(-\frac{1}{2}\frac{\pp^2}{\pp \alpha^2}\right)\alpha^n$ bridges to the Hermite-polynomials $\text{He}_n(\alpha) $ and ultimately to the Gram-Charlier expansion: Approximating $g(\alpha)$ as a polynomial would automatically lead to a polynomial partition function in the spirit of \autoref{flow_expansion}.}
This yields by usage of the well known Faà di Bruno formula a complicated but manageable expression. With the help of Bell polynomials, this falls back to 
\begin{align}\label{exp. val. end result NF}
	\langle \theta^m \rangle 
	= \sum_{k=0}^{\infty} \sum_{i=0}^{2k} \frac{h^{(i)}}{2^k k!}(\theta(0)) B_{2k,i}(\theta^{(1)}(0),\ldots,\theta^{(2k+1-i)}(0)),
\end{align}
with $h(x) = x^m$. Although the series involving the Bell polynomials can be generalised to higher dimensions, it is numerically very expensive to compute and thus will not be used. For our normalising flow architecture, it is better to compute the series in \autoref{equ:laplace_1d} by iterative usage of the implemented \texttt{autograd} functionality of PyTorch. In higher dimensions and for general expectation values for probability distributions, one finds with the help of \autoref{flow_expansion}
\begin{align}
    \label{flow_expansion_lapace}
	\langle \theta^{\gamma_1} \ldots \theta^{\gamma_m} \rangle 
	= \sum_{k=0}^{\infty} \frac{1}{2^k k!} \left(\delta^{\rho \sigma} \frac{\partial}{\partial\alpha^\rho}\frac{\partial}{\partial\alpha^\sigma} \right)^k \theta^{\gamma_1}(\alpha) \cdots \theta^{\gamma_m}(\alpha),
\end{align}
again taken at $\alpha = 0$. This equation will be used for moment computations in the following chapters.

\section{Application to supernova data}\label{sect_cosmo}
We are able to use the entropy calculation as well as the flow expansion for a more complex and topical example of supernova data allowing to derive constraints on certain parameters - a common setup, see for example \citet{riess1998observational} or \citet{herzog2023partition}. The parameters of interest are the matter density $\Omega_m$ and the dark energy equation of state. It was for example in \citet{modelselection} shown that the most likely model is a constant parameter $w_0$. Thus, we try to recover estimates for those parameters using the flow expansion via a series of derivatives after calculating the information entropy of this setup using the normalising flow trained with \texttt{FrEIA} \citep{freia}.

\begin{figure}[H]
    \centering
    \includegraphics[width=0.47\textwidth]{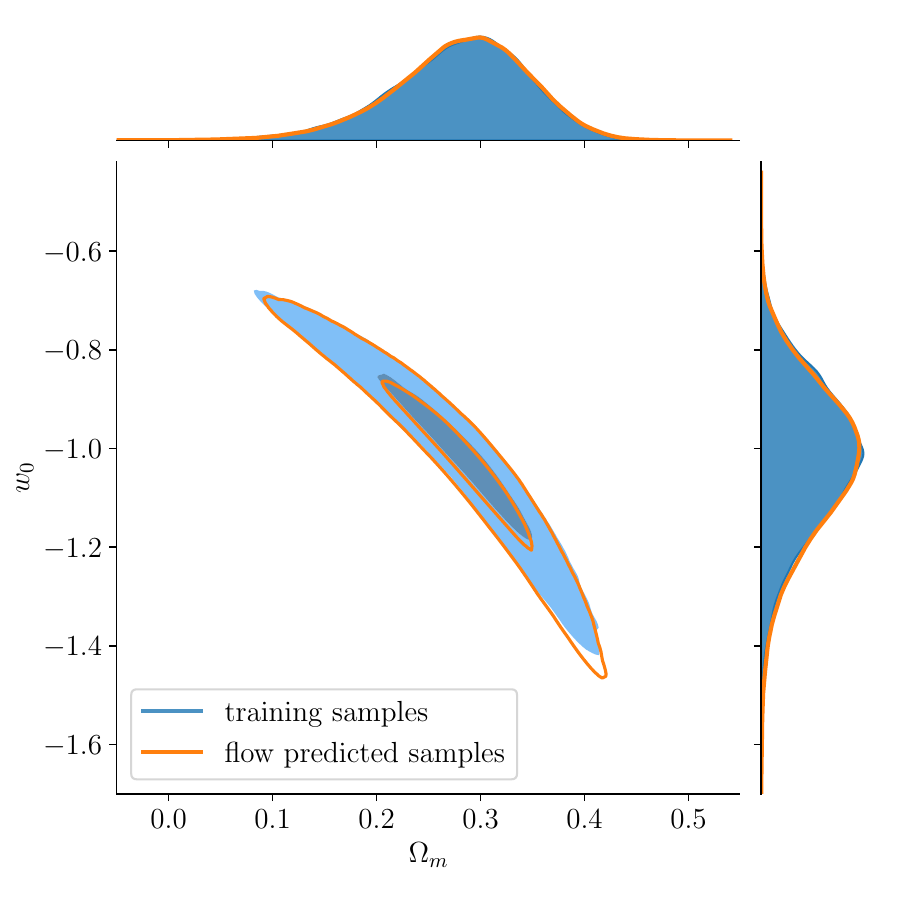}
    \caption{The normalising flow (orange) reproduces the posterior samples (blue) of the supernova Ia example, thus the two distribution as well as their marginals match.}
    \label{fig:supernova_comp}
\end{figure}

The Union2.1 data set \citep{suzuki2012hubble, kowalski2008improved, amanullah2010spectra} of supernovae of type Ia is used and contains 580 measurements. The distance modulus $y$ is defined by the difference between the apparent magnitude $m$ and the absolute magnitude $M$, and can be related to the luminosity distance $d_L(a|\Omega_m, w_0)$ as
\begin{equation}
    \label{eq:distance_modulus}
    y = m - M = 5 \log_{10}(d_L(a|\Omega_m, w_0))+10 \,.
\end{equation}
The luminosity distance is given by
\begin{equation}
    \label{eq:lumi_distance}
    d_L(a|\Omega_m, w_0)= \frac{c}{a} \int_a^1\dd a^\prime \frac{1}{a^{\prime 2} H(a^\prime,\Omega_m, w_0)}
\end{equation}
with $H(a|\Omega_m, w_0)$ denoting the Hubble function $H(a,\Omega_m, w_0)$ which is of the following expression for a constant, in terms of the scale factor $a$, equation of state parameter $w_0$:
\begin{equation}
    \label{eq:hubble_function}
    \frac{H(a|\Omega_m, w_0)^2}{H_0^2} = \frac{\Omega_m}{a^3}+\frac{1-\Omega_m}{a^{3(1+w_0)}}
\end{equation}
with the Hubble parameter fixed at $H_0 = 70\,\text{km/s/Mpc}$. An analytical solution to \autoref{eq:lumi_distance}, expressed by means of a hypergeometric function, exists \citep{rafael_paper}.

Getting an analytical solution for the distance modulus $\tilde{y}(a_i | \Omega_m, w_0)$ allows under the assumption of Gaussian errors $\sigma_i$ for each measurement to define the likelihood to be
\begin{equation}
    \likeli (y | \Omega_m, w_0) \propto \exp\left(-\frac{1}{2}\sum_{i}\left(\frac{y_i - \tilde{y}(a_i | \Omega_m, w_0)}{\sigma_i}\right)^2\right)\,.
\end{equation}
Physically motivated, we choose a uniform prior for $\Omega_m$ and $w_0$.
 This allows to compare our results for example to \citet{partition_info}, which is especially interesting for the calculated entropy. We used the package \texttt{PyMultiNest} \citep{pymutltinest, feroz2009multinest} to obtain a value for the Bayesian evidence $p(y)$ as well as the \texttt{emcee} package \citep[c.p.][]{emcee_citation} to get posterior samples, which are together with their normalising flow reconstruction presented in \autoref{fig:supernova_comp}.

 Making use of \autoref{equ:exp_val_approx}, the information entropy at unit temperature could be calculated via a sample estimate as
\begin{equation}
    \label{eq:entropy_sample}
    S_\mathrm{sample} = - \frac{1}{N} \sum_{i=1}^N \ln p(\theta^i | y) \,
\end{equation}
with $N$ being the number of samples. This requires a distributional form of the posterior $p(\theta | y)$, which is not analytically known. For comparison, we estimate $p(\theta^i |y)$ on a grid (effectively forming a histogram) and by kernel density estimation (KDE), using the \texttt{SciPy} package. Those approximations are not needed when working with the normalising flow: It provides a function $f(\theta)$ that maps the samples to a standard normal distribution $p(\alpha)$, with known analytical form. Thus, by sampling from a normal distribution and using the trained normalising flow, the entropy can be computed via
\begin{equation}
    \label{eq:entropy_flow}
    S_{\mathrm{flow}} = - \frac{1}{N} \sum_{i=1}^N \left( \ln p(\alpha^i) + \ln | \det \mathrm{D} f(\alpha^i) | \right) \,.
\end{equation}
The advantages of calculating the entropy using a transformation to a known (Gaussian) probability distribution were shown in \citet{ao2022entropy}, for instance. \texttt{FrEIA} provides the Jacobian of the transformation so that no approximations - except for using the normalising flow to obtain the map $f(\alpha)$ are needed. \autoref{tab:supernova_entropies} presents the values for the information entropy of the supernova posterior. All values and for most the one obtained via the transformation using the normalising flow match within their errors.

\begin{table}[H]
    \centering
    \begin{tabular}{@{}llll@{}}
        \toprule
             & histogram & KDE & flow \\ \midrule
             $S$  & $-3.359 \pm 0.004$ & $ -3.350 \pm 0.004$ & $-3.359 \pm 0.019$\\  \bottomrule
    \end{tabular}
    \caption{Information entropy $S$ for supernova - calculations via histogram \citep[as also done in][]{partition_info}, kernel density estimate (KDE) and via normalising flow (flow). All values match within the given errors.}
    \label{tab:supernova_entropies}
\end{table}

The aim of most statistical analysis is to derive constraints on certain parameters - in our case on the matter density $\Omega_m$ and the dark energy equation of state parameter $w_0$. We can do so by finding the maximum posterior which coincides with the mean of the posterior samples. The mean, i.e.\ the first moment, can be calculated via the samples or more interestingly using the flow expansion via a series of derivatives. The results are shown in \autoref{tab:supernova_moments_mf}. The values match not only each other, but also the usually obtained values for the supernova data as for example in \citet{suzuki2012hubble}.

\begin{table}[H]
    \centering
    \begin{tabular}{@{}llll@{}}
        \toprule
             & $\bra \Omega_m \ket$ & $\bra w_0 \ket$ & Cov ($\Omega_m \cdot  w_0$) \\ \midrule
        sampling  & $0.2759 \pm 0.0006$ & $-1.0143 \pm 0.0015$ & $-0.0101 \pm 0.0004$\\ 
        expansion  & $0.2777 \pm 0.0017$ & $ -1.0156 \pm 0.0021$ & $-0.0091 \pm 0.0019$\\ 
        posterior & $0.2759 \pm 0.0003$ & $-1.0138 \pm 0.0008$ & $-0.0095 \pm 0.0001$\\ 
        \bottomrule
    \end{tabular}
    \caption{Comparison of moments derived from sampling as described in \autoref{equ:exp_val_approx} and via the flow expansion (\autoref{flow_expansion_lapace}) for the supernova data. Additionally, the sample estimates on the \texttt{emcee} posterior samples are given. Not only agree the values nicely, but also the results match the usually obtained ones \citep[e.g.][]{suzuki2012hubble}.}
    \label{tab:supernova_moments_mf}
\end{table}

This shows that the via the flow expansion statistical analysis of real world data is possible and agrees within its errors with our expectation. Concerning numerical performance, the accuracy of the flow expansion for the mean and variance is a little worse than the sampling method, but still in the same order of magnitude. Both methods are roughly equally fast. The goal of the flow expansion is not to be numerically more performant, but to offer analytical improvements for statistical partition functions. The present limitation comes from the fact that higher derivatives of normalising flows are not possible yet and smoothness can be improved. Nevertheless, also more basic methods as information entropy calculation become easier as one is able to insert known analytical distributions and just uses the Jacobian of the transformation performed by the neural network.

\section{Geometry of the Normalising Flow}\label{sect_geo}
In this chapter, we will investigate how the invertible neural network maps Gaussian to non-Gaussian distributions by analysing the geometry of the nonlinear mapping.

\begin{figure}[H]
    \centering
    \includegraphics[width=0.47\textwidth]{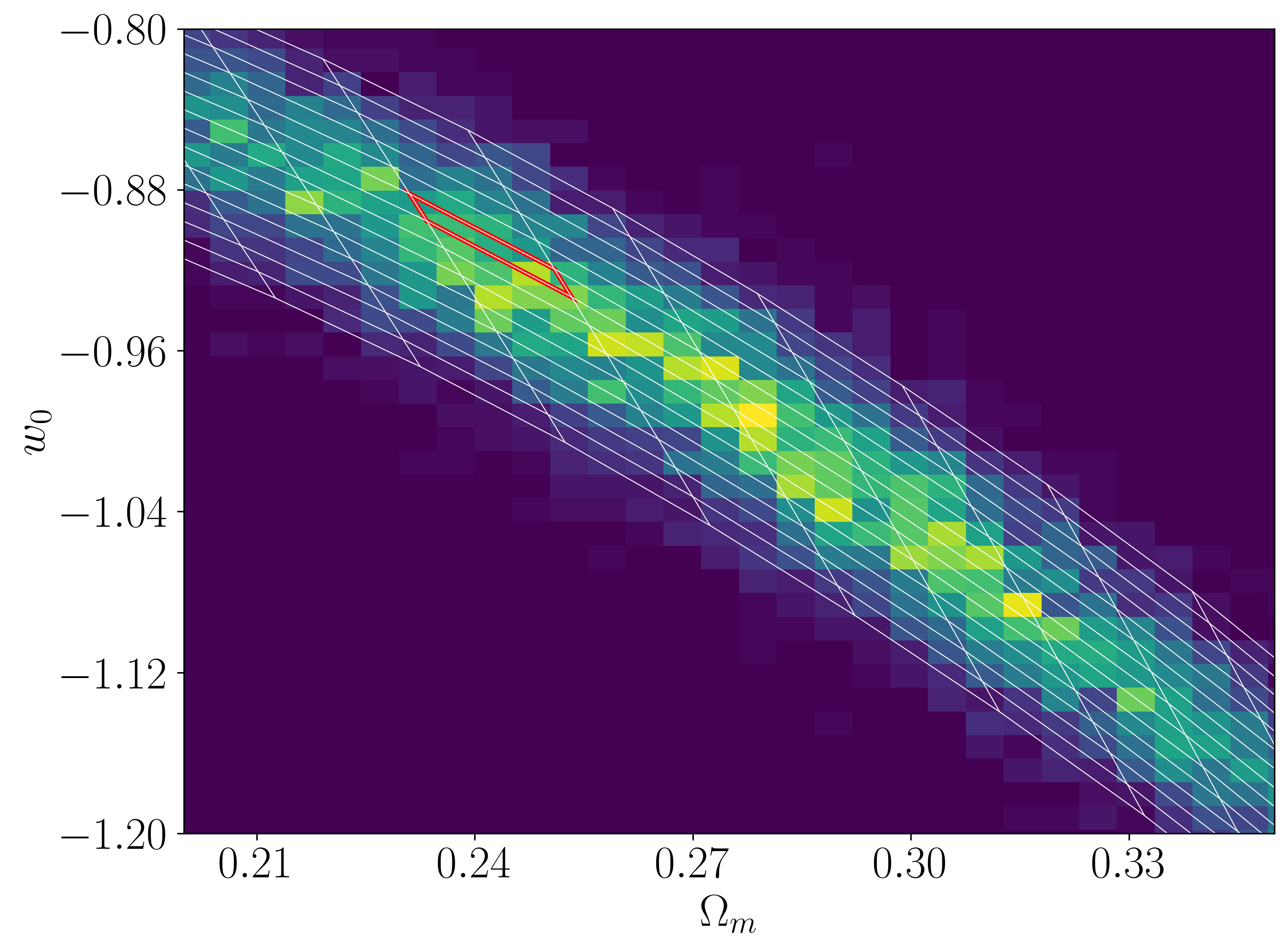}
    \caption{Geometric visualisation of the transformation induced by the normalising flow  - zoom in on the maximum a posteriori region. The specific cell marked in red is analysed in more detail. The shading in the background represents the posterior probability.}
    \label{fig:Geometrical Normalising Flow}
\end{figure}

In \autoref{fig:Geometrical Normalising Flow}, one can see the transformation of a Cartesian, rectilinear grid under the normalising flow - constructed for the posterior distribution for $\Omega_m$ and $w_0$ in the supernova-example, for a zoom-in centered on the maximum of the posterior distribution. Borrowing an idea from gravitational lensing, we decompose the Jacobian matrix in terms of a basis constructed from the Pauli-matrices
\begin{align}\label{eq:Decomposition matrix}
	Jf = 
	\kappa \begin{pmatrix} 
		1 & 0 \\ 
		0 & 1 
	\end{pmatrix} 
	+ \gamma_1 \begin{pmatrix}
		1 & 1 \\
		0 & -1 
	\end{pmatrix}
	+ \gamma_2 \begin{pmatrix}
		0 & 1 \\
		1 & 0
	\end{pmatrix}
	+ \omega \begin{pmatrix}
		0 & 1 \\
		-1 & 0
	\end{pmatrix} \,,
\end{align}
it is possible to quantify the amount of isotropic change of size $\kappa$, anisotropic shearing $\gamma_1$, $\gamma_2$ and rotation $\omega$ that any grid cell undergoes while being mapped by the flow. Because the mapping is nonlinear and naturally position dependent, we focus on the red grid cell in \autoref{fig:Geometrical Normalising Flow} as an example: 
\begin{align}
	Jf = \begin{pmatrix} 
		-0.0398 & -0.1233  \\
		0.0216 & -0.0620  
	\end{pmatrix}\,.
\end{align}

Calculating the coefficients introduced in \autoref{eq:Decomposition matrix} for this matrix and rounding up to  four digits, yields $\kappa = -0.0509$, $\gamma_1 = 0.0111$, $\gamma_2 = -0.0564$, and $\omega = -0.0780$. Additionally, we note that the flow conserves orientation of the coordinate frames. The areas of cells in parameter spaces in $\theta$ and $\alpha$ are naturally related by $\kappa^2$, as verified in \autoref{tab:vol_geometry_decomposition}.

\begin{table}[H]
	\centering
	\begin{tabular}{@{}llll@{}}
		\toprule
		$\log \det Jf$ & $ V(\Omega_m,w_0)$ & $V(\alpha_1,\alpha_2)$ & $\kappa^2 \cdot V(\alpha_1,\alpha_2)$\\ \midrule
		$-6.0901$ & $0.0002$ & $0.0816$ & $0.0002$ \\ \bottomrule
	\end{tabular}
	\caption{Belonging to the red marked cell in \autoref{fig:Geometrical Normalising Flow} - the $\log$ determinant of the Jacobian as well as the corresponding volumes for the cell of the true posterior and the standard normal are given. Multiplying the latter by $\kappa^n$ yields for two dimensions ($n=2$) the volume in terms of the posterior.}
	\label{tab:vol_geometry_decomposition}
\end{table}

\section{State variables $T$ and $J_\gamma$ of Bayesian partitions}\label{sect_temp_J}
The normalising flow allows to explore the thermodynamic quantities of the partition function defined in \autoref{eq:canZ} easily as sampling from a Gaussian and transforming back is numerically fast and convenient. Combining the partition function $Z[T, J]$ in \autoref{flow_expansion} with the statistical estimate of the expectation value in \autoref{equ:exp_val_approx}, one obtains
\begin{align}
    \label{flow_partition_sum}
    Z[T, J] 
    &\approx \frac{p(y)^{\frac{1}{T}}}{N} \sum_{i=1}^N \left( \frac{ p(\alpha^i)}{\left | \det \mathrm{D} f(\alpha^i) \right |} \right)^{\frac{1}{T} - 1} \exp \left( \frac{J_{\gamma}\theta^{\gamma} (\alpha^i)}{T} \right) 
\end{align}
with $p(\alpha)$ being the standard normal distribution. Note, that the normalisation $\mathcal{N}(T)$ has been inserted taking care of the fact, that the posterior and not the product of likelihood and prior was learned. This procedure is necessary as the analytical from of $p(\alpha)$ is known and thus, one can perform the estimate with samples $\alpha^i$ drawn from a standard Gaussian.

\begin{figure}[H]
    \centering
    \includegraphics[width=0.43\textwidth]{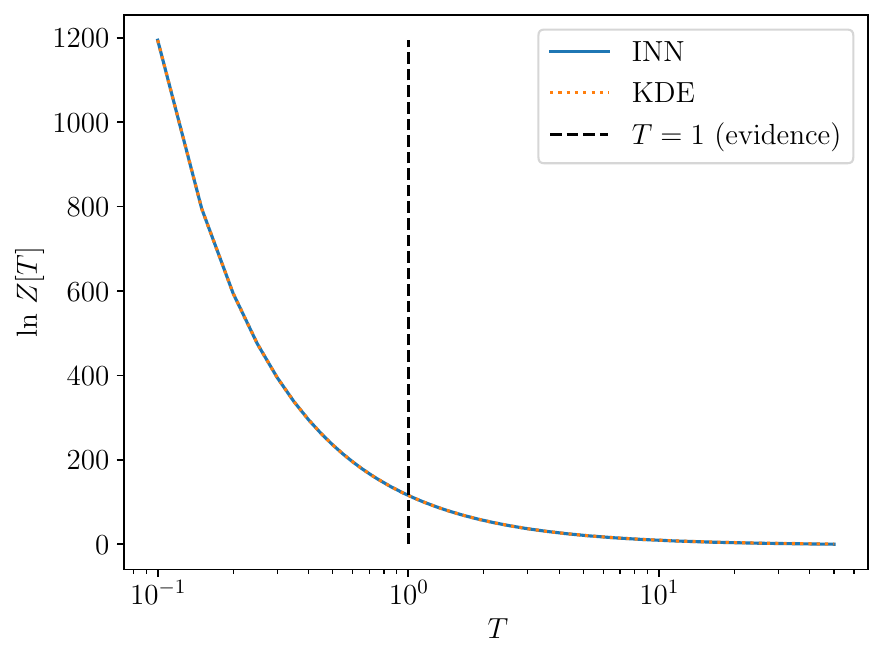}
    \caption{Plot of the partition function $\ln Z$ as a function of temperature $T$ with $J=0$ - comparing the flow (blue) result to KDE (orange) estimate as in \citet{partition_info}. The two results agree perfectly.}
    \label{fig:lnZt}
\end{figure}

For example, \autoref{fig:lnZt} shows the logarithm of the partition function as a function of temperature, similarly to \citet{partition_info}, where a KDE approach (displayed in orange) was used. For simplicity, $J$ is set to zero. One can nicely see that both approaches agree almost perfectly. As expected, the value of $\ln Z[T]$ saturates for high temperatures $T$.

\begin{figure}[H]
    \centering
    \includegraphics[width=0.45\textwidth]{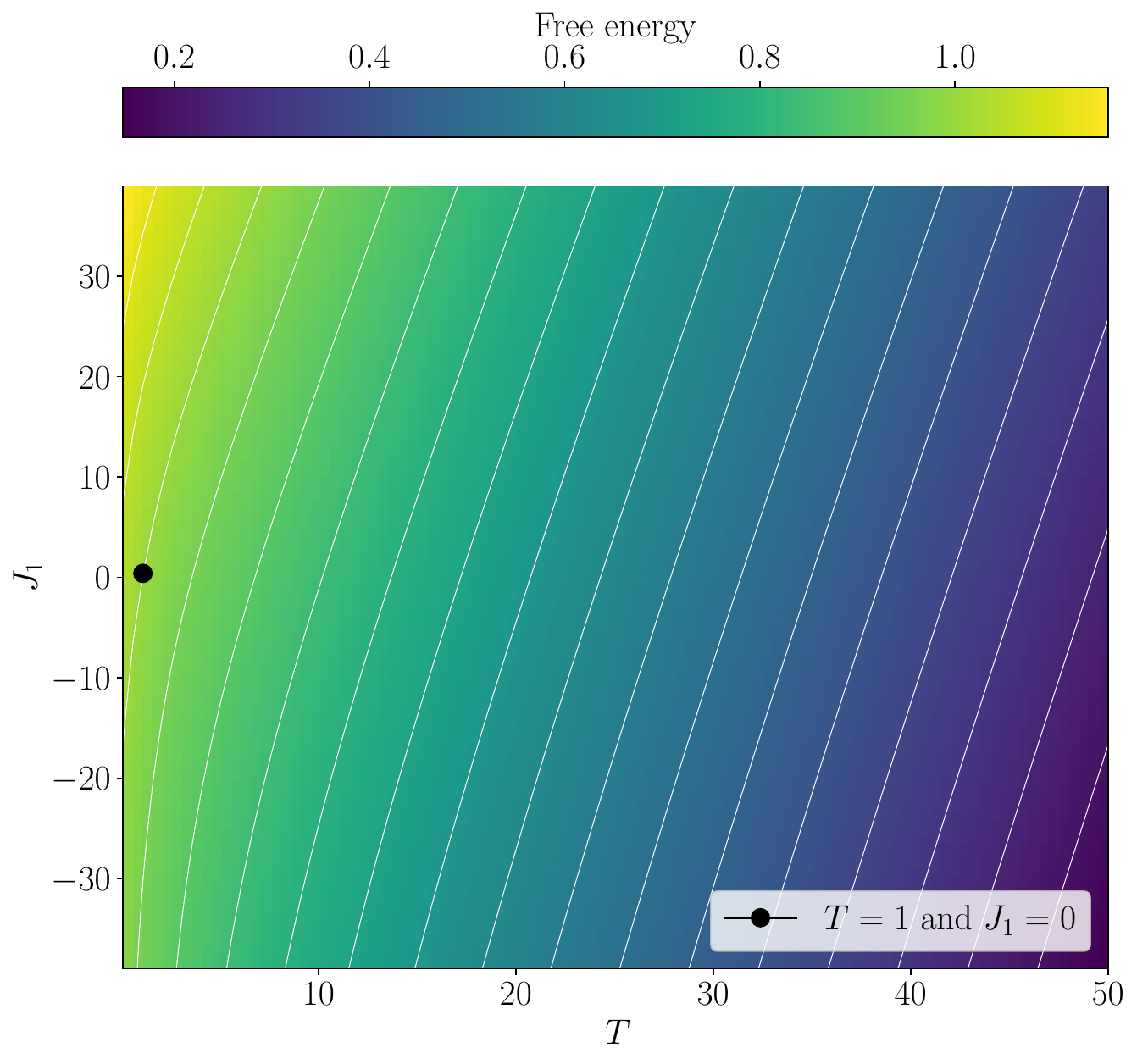}
    \caption{Helmholtz free energy (\autoref{equ:free_energy}) as a function of temperature $T$ and $J_1$ while $J_2 = 0$, normalised to its value at $T=1$ and $J_\gamma = 0$ for $\gamma \in \{1,2\} $.}
    \label{fig:free_enegergy_TJ}
\end{figure}

\begin{figure}[H]
    \centering
    \includegraphics[width=0.45\textwidth]{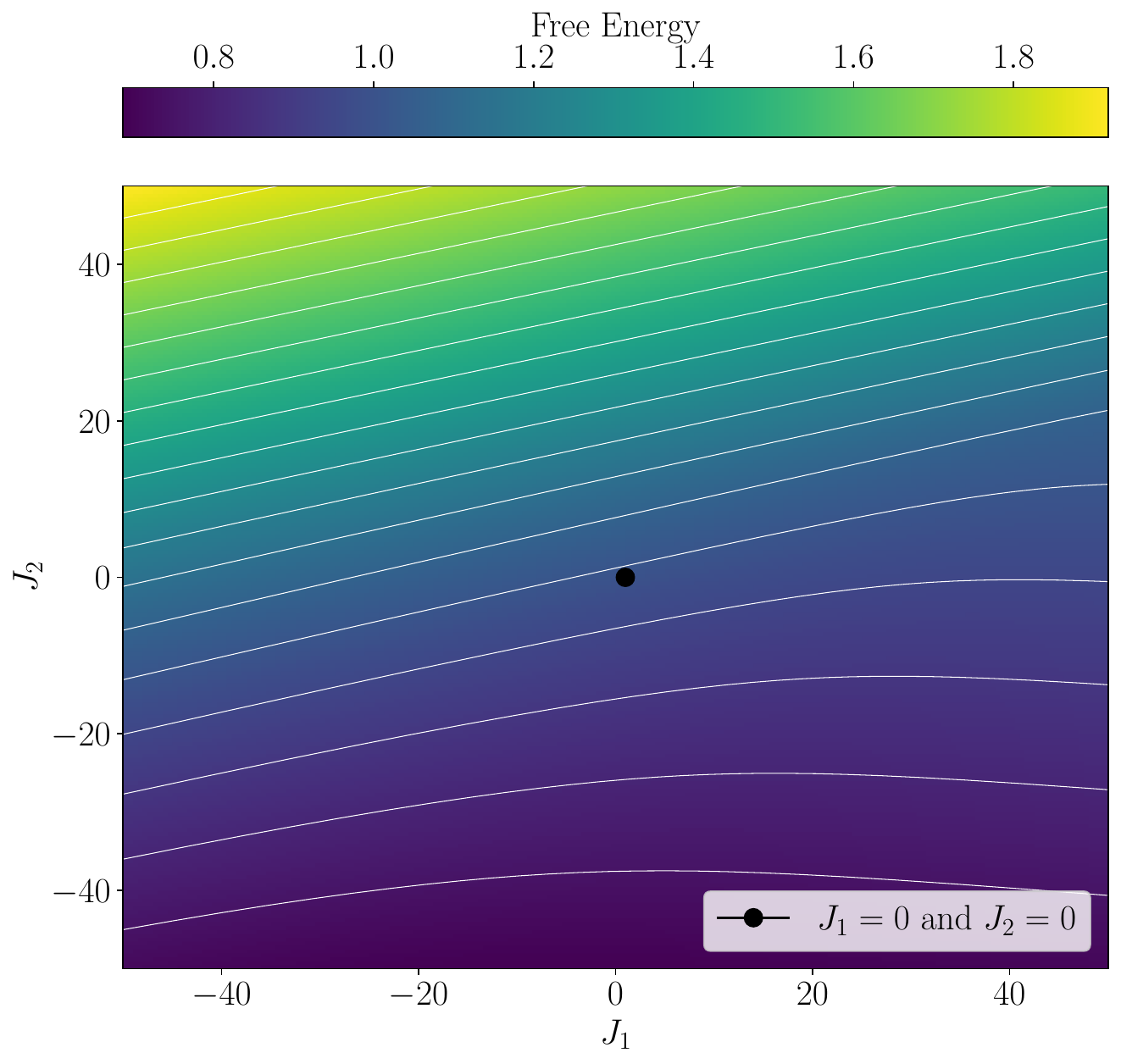}
    \caption{Helmholtz free energy (\autoref{equ:free_energy}) as a function of $J_1$ and $J_2$ at temperature $T=1$ - normalised to its value at $J_\gamma = 0$ for $\gamma \in \{1,2\}$.}
    \label{fig:free_energy_JJ}
\end{figure}

As shown in \autoref{fig:free_enegergy_TJ} and \autoref{fig:free_energy_JJ}, we can also use the normalising flow to visualise the free energy defined in \autoref{equ:free_energy} as a function of temperature $T$ and the hyperparameters $J_1$ and $J_2$. Within these figures it is normalised to the fiducial value at $F(T=1, J_1 =0, J_2=0)$ which corresponds to the Bayesian evidence.

In both plots, the isocontours reveal that there exist certain choices of $\{T, J_1\}$ and $\{J_1, J_2\}$ such that the Helmholtz free energy is equal to the Bayesian evidence at $T=1, J_1 =0$ and $J_2 = 0$. Thus, one can compensate for temperature changes by adjusting the state variables $J_1$ and $J_2$ accordingly, with possible advantages in sampling.

\section{summary and discussion}\label{sect_summary}
The subject of this paper was a hybrid approach to Bayesian inference with non-Gaussian distributions, combining normalising flows as a numerical machine learning method with partition functions as an analytical method for computing cumulants and entropies. Normalising flows construct a differentiable and invertible mapping between an ideal, Gaussian distribution and a non-Gaussian distribution, which allows insights into non-Gaussianity of the posteriors.

\begin{enumerate}[(i)]
\item{Normalising flows are well-working numerical techniques for the evaluation of Bayesian evidences for non-Gaussian distributions \citep{2024arXiv240412294S}. Extending the expression for the Bayesian evidence to a Bayesian partition sum by the inclusion of a sampling temperature $T$ and a generating variable $J_\gamma$ does not influence the numerics of the normalising flow: Computations of partition functions across the space spanned by $T$ and $J_\gamma$ are well possible and allow the exploration of this space.}
\item{Computations of the information entropy become particularly straightforward. The change of variables formula allows to rewrite the information entropy in terms of the standard normal distribution allowing to perform the sample estimate for an analytically known distribution. The results agree perfectly with the values obtained by kernel density estimates, and by those from derivatives of the Helmholtz free energy $F(T,J_\gamma)$ with respect to temperature $T$ as in \cite{partition_info}.}
\item{By suitable differentiation of the logarithmic partition sums, the moments of the posterior distribution can be obtained. This involves higher order derivatives of the learned normalising flow, which are numerically problematic; \autoref{tab:supernova_moments_mf_complete_omega_m} and \autoref{tab:supernova_moments_mf_complete_w0} show that autodifferentiability becomes unreliable beyond the second moments. Mean, variance and covariance of the posterior distribution are obtained almost perfectly, though, and algorithmic advances may remedy the issue.}
\item{The mapping constructed by the normalising flow can in two dimensions be interpreted geometrically and decomposed in terms of shearing, rotation and scaling of volume elements. The action of the determinant of the Jacobian of the variable change modulates the initially Gaussian distribution onto the required functional shape, complementing \citep{schafer_describing_2016}.}
\end{enumerate}

In summary, we report on an integration of three concepts: Bayesian inference, normalising flows and partition functions for improving sampling, the analytical characterisation of non-Gaussian posterior distributions and the derivation of quantities like information entropies. We intend to improve further differentiability and optimise network layouts for that purpose. As an alternative, a physics-informed neural network can learn the partition function $Z[T,J_\gamma]$ in its dependence on the state variables $T$ and $J_\gamma$ directly, and yield thermodynamic relations through differentiation.

\begin{table}[H]
    \centering
    \begin{tabular}{@{}lrrr@{}}
        \toprule & sampling & expansion & posterior (emcee) \\ \midrule
        $\bra \Omega_m \ket$ & $0.2759 \pm 0.0006$ & $0.2777 \pm 0.0017$ & $0.2759 \pm 0.0003$ \\
        Var($\Omega_m$) & $0.00463 \pm 0.00014$ & $0.0043 \pm 0.0003$ & $0.00440 \pm 0.00002$ \\
        $s^\gamma$ & $-0.55 \pm 0.07$ & $-0.9 \pm 0.3$ & $-0.531 \pm 0.013$ \\
        $\kappa^\gamma$ & $0.77 \pm 0.12$ & $6.2 \pm 1.5$ & $0.55 \pm 0.05$ \\ \bottomrule
    \end{tabular}
    \caption{Supernova - evaluation of mean, variance, skewness, and kurtosis for $\Omega_m$ via the flow expansion compared to sampling estimates via the flow learned distribution (sampling) and the \textsl{emcee} posterior, which can be regarded as ground truth.}
    \label{tab:supernova_moments_mf_complete_omega_m}
\end{table}

\begin{table}[H]
    \centering
    \begin{tabular}{@{}lrrr@{}}
        \toprule & sampling & expansion & posterior (emcee)\\ \midrule
        $\bra w_0 \ket$ & $-1.0143 \pm 0.0015$ & $-1.0156 \pm 0.0021$ & $-1.0138 \pm 0.0008$ \\
        Var($w_0$) & $0.0238 \pm 0.0014$ & $0.0216 \pm 0.0019$ & $0.02231 \pm 0.00014$ \\
        $s^\gamma$ & $-0.38 \pm 0.09$ & $-0.56 \pm 0.14$ & $-0.318 \pm 0.011$ \\
        $\kappa^\gamma$ & $0.64 \pm 0.27$ & $5.2 \pm 1.3$ & $0.21 \pm 0.04$ \\ \bottomrule
    \end{tabular}
    \caption{Supernova - evaluation of mean, variance, skewness, and kurtosis for $w_0$ via the flow expansion compared to sampling estimates via the flow learned distribution (sampling) and the \textsl{emcee} posterior, that can be seen as ground truth.}
    \label{tab:supernova_moments_mf_complete_w0}
\end{table}

\section*{Acknowledgements}
\paragraph{Funding information}
We acknowledge the usage of the AI-clusters {\em Tom} and {\em Jerry} funded by the Field of Focus 2 of Heidelberg University.

\paragraph{Thanks}
We are grateful to Lennart Röver, Benedikt Schosser, Rebecca Maria Kuntz, Maximilian Philipp Herzog and Heinrich von Campe for insightful discussions, and to Ulli K{\"o}the and Hans Olischl{\"a}ger for providing support on \texttt{FrEIA}.

\paragraph{Data availability}
Our Python implementation of the code computing cumulants and entropies from the normalising flow is available on \href{https://github.com/cosmostatistics/partition-nf-expansion}{GitHub}.

\appendix

\section{Toy model}\label{sect_toy_model}
\begin{figure}[H]
    \centering
    \includegraphics[width=0.47\textwidth]{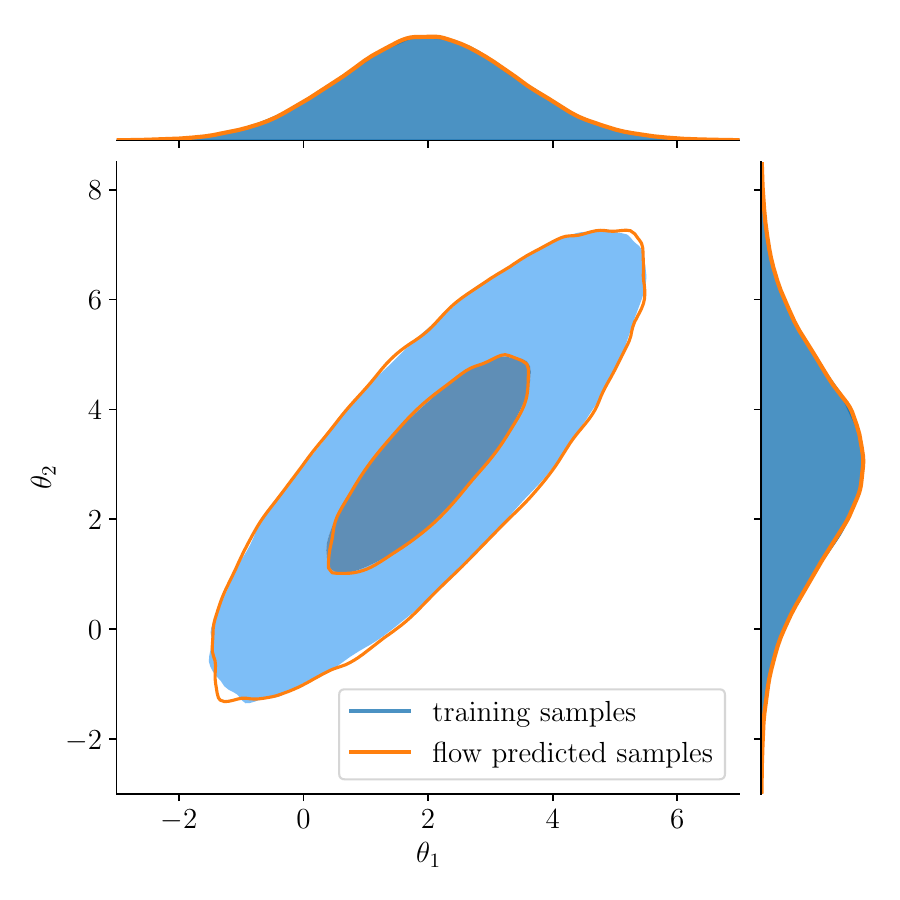}
    \caption{The inverted normalising flow (orange) reproduces nicely the posterior samples (blue) of the toy model.}
    \label{fig:toy_model_comp}
\end{figure}

For verification purposes, we use the normalising flow method with a Gaussian toy model, for ensuring that the moment and entropy calculations as well as the flow expansion via a series of derivatives is valid. At the same time, we optimise the parameters of \texttt{FrEIA}. The toy model is defined by a Gaussian distribution with mean $\mu$ and covariance $C$ given by
\begin{equation*}
    \mu_{\text{toy}} = \left( \begin{array}{c} 2 \\ 3 \end{array} \right) 
    \quad \text{and} \quad 
    C_{\text{toy}} = \left( \begin{array}{cc} 2 & 2 \\ 2 & 3 \end{array} \right) \,.
\end{equation*}
The corresponding training samples and its reconstruction with the normalising flow using the software package \texttt{FrEIA} \citep{freia} are shown in \autoref{fig:toy_model_comp}.

Throughout this paper, the sequential invertible neural network (\texttt{SequenceINN}) architecture with the so called \texttt{AllInOneBlock} of \texttt{FrEIA} are used. The latter combines affine coupling, permutations and a global affine transformation, for more details on the architecture refer to \citet{freia} or \citep{freia_background}, introducing the theory behind \texttt{FrEIA}. The Adam optimiser is chosen. As the toy model already is a Gaussian distribution, just shifted by its mean $\mu_{\text{toy}}$ and scaled by its covariance matrix $C_{\text{toy}}$, the required complexity is quite low. Thus, one layer and a subnet width of 64 (used in the \texttt{AllInOneBlock}) are sufficient. In contrast, for the supernova application more complexity is needed and thus two broader layers with a subnet width of 128 are used. 

The toy model offers the possibility to calculate the information entropy analytically. For a two-dimensional normal distribution it is given by
\begin{equation}
    S_{2d \, \mathrm{Gaussian}} = \ln 2\pi + \frac{1}{2} \ln \det C + 1 \,.
\end{equation}
Again, the entropy is also calculated using the normalising flow and \autoref{eq:entropy_flow}. \autoref{tab:toy_model_entropies} compares those values to estimates obtained via a grid (effectively forming  histogram) and by kernel density estimation (KDE).

\begin{table}[H]
    \centering
    \begin{tabular}{@{}lllll@{}}
        \toprule
               & analytical & histogram & KDE & flow \\ \midrule
        $S$  & $3.1845$ & $3.1835 \pm 0.0008$ & $3.2031 \pm 0.0008$ & $3.1833 \pm 0.0019$\\  \bottomrule
    \end{tabular}
    \caption{Information entropy $S$ for the toy model - analytical value (analytical) and calculations via histogram (histogram), kernel density estimate (KDE) and through normalising flows (flow). The flow obtained value perfectly coincides with the analytical value.}
    \label{tab:toy_model_entropies}
\end{table}

All errors are estimated by repeating the training procedure of the normalising flow on ten different data sets, drawn from the same ground truth. This comparison allows at least two conclusions: For most, the flow obtained entropy value perfectly coincides with the analytical one proving the method. Secondly, the KDE calculated entropy does not match the ground truth within its error which quite likely is underestimated because it comes from repetition and lacks information about the system error of a kernel density estimate. As being an additional step, there is more room for error.

\begin{table}[H]
    \centering
    \begin{tabular}{@{}lllll@{}}
        \toprule & $\theta_1$ (gt) & $\theta_1$ (expansion) & $\theta_2$ (gt) & $\theta_2$ (expansion) \\ \midrule
        $\bra \theta^\gamma \ket$ & $2$ & $2.0003 \pm 0.0010$ & $3$ & $2.92 \pm 0.08$\\ 
        Var($\theta^\gamma$) & $2$ & $2.0004 \pm 0.0023$ & $3$ & $3.12 \pm 0.12$ \\ 
        $s^\gamma$ & $0$ & $(5 \pm 7) \times 10^{-16}$ & $0$ & $0.13 \pm 0.18$\\ 
        $\kappa^\gamma$ & $3$ & $3 \pm 1.6 \times 10^{-15}$ & $3$ & $3.0 \pm 0.3$\\ 
        Cov($\theta_1 \theta_2$) & $2$ & $2.09 \pm 0.09$ & - & - \\ \bottomrule
    \end{tabular}
    \caption{Toy model - evaluation of mean, variance, skewness and kurtosis for each direction $\theta_1$ and $\theta_2$ via the flow expansion compared to the ground truth (gt) of the generated data. All values match the ground truth within the errors.}
    \label{tab:toy_model_moments_mf}
\end{table}

The flow expansion with a series of derivatives defined in \autoref{flow_expansion} is applied to the toy model to calculate in each dimension the mean, variance, skewness as well as kurtosis and the covariance of both dimensions. The results are shown in \autoref{tab:toy_model_moments_mf}: First, the first dimension is not only closer to the correct value, but also its error is smaller than obtained for the second dimension. This is related to the procedure how the normalising flow is constructed within \texttt{FrEIA}. There, the second dimension uses the output from the first dimension \citep{freia_background} which leads to a propagation of error. One could avoid that problem by training the normalising flow a second time and swapping the input dimensions. Additionally, it is apparent that the network has learned a transformation of a Gaussian distribution to another Gaussian distribution as the values for skewness and kurtosis are especially in the first dimension very precise and exactly what is expected for a Gaussian distribution - independently of its mean and covariance: The normalising flow has to be able to reproduce a principal value decomposition which is a mere linear transform between the random variables. The obtained value for the covariance of $\theta_1$ and $\theta_2$ still agrees within its error with the ground truth, but the larger error and deviation originates from the usage of the second dimension in the calculation.

Throughout this paper, we used the flow expansion up to fourth order, i.e. terminating the series defined in \autoref{flow_expansion_lapace} at $k=4$. This is due to the fact that the higher derivatives of the network become more and more difficult, leading to divergences. This is deeply connected to the software architecture. Even for the compulsory choice of smooth activation functions, other aspects of the algorithm are not smooth and thus do not allow for a high number of derivatives. For example, the performed permutation in the \texttt{AllInOneBlock} is found to be very important for training, but impacts derivatives negatively. Normalising flows and as an example \texttt{FrEIA} are capable of learning far more complicated distributions, but for a higher complexity more layers are need. From our experience, this results in issues with higher-order derivatives of the normalising flow. But if given a smooth normalising flow which is sufficiently often differentiable, one would be able to apply the flow expansion to any order and to an arbitrary complex distribution. There are first attempts to create smooth normalising flows like for example \citet{kohler2021smooth}, and it would be interesting to evaluate their performance beyond second order.

\vspace{5mm}

\bibliography{references}

\end{document}